\documentclass[aps,prl,10pt,twocolumn,amsmath,amssymb,notitlepage,superscriptaddress]{revtex4-1}
\usepackage{tikz, graphicx, xcolor, array, times, mathtools, qcircuit}
\usepackage[T1]{fontenc}

\newcolumntype{C}{>{$}c<{$}}
\AtBeginDocument{
\heavyrulewidth=.08em
\lightrulewidth=.05em
\cmidrulewidth=.03em
\belowrulesep=.65ex
\belowbottomsep=0pt
\aboverulesep=.4ex
\abovetopsep=0pt
\cmidrulesep=\doublerulesep
\cmidrulekern=.5em
\defaultaddspace=.5em
\tabcolsep=7pt
}
\usepackage{booktabs}

\newcommand{\ket}[1]{|#1\rangle}
\newcommand{\bra}[1]{\langle #1|}
\newcommand{\ftt}{[4,2,2]}

\usepackage{hyperref}
\definecolor{darkblue}{RGB}{0,0,127} % choose dark colors for high contrast
\definecolor{darkgreen}{RGB}{0,150,0}
\hypersetup{breaklinks, colorlinks, linkcolor=darkblue, citecolor=darkgreen, filecolor=red, urlcolor=darkblue, pdftitle={Fault tolerance in the IBM Q Experience}, pdfauthor={Robin Harper and Steven Flammia}}% add a title to the metadata

\def\equationautorefname~#1\null{Eq.~(#1)\null}

\begin{document}

\title{Fault-Tolerant Logical Gates in the IBM Quantum Experience}
\author{Robin Harper}
\affiliation{Centre for Engineered Quantum Systems, School of Physics, The 
University of Sydney, Sydney, Australia}
\author{Steven T. Flammia}
\affiliation{Centre for Engineered Quantum Systems, School of Physics, The 
University of Sydney, Sydney, Australia}
\affiliation{Yale Quantum Institute, Yale University, New Haven, CT 06520, USA}
\date{\today}

\begin{abstract}
Quantum computers will require encoding of quantum information to protect them from noise. 
Fault-tolerant quantum computing architectures illustrate how this might be done, but have not yet shown a conclusive practical advantage. 
Here we demonstrate that a small but useful error detecting code improves the fidelity of the fault-tolerant gates implemented in the code space as compared to the fidelity of physically equivalent gates implemented on physical qubits. 
By running a randomized benchmarking protocol in the logical code space of the \ftt~code, we observe an order of magnitude improvement in the infidelity of the gates, with the two-qubit infidelity dropping from $5.8(2)\%$ to $0.60(3)\%$. 
Our results are consistent with fault-tolerance theory and conclusively demonstrate the benefit of carrying out computation in a code space that can detect errors. 
Although the fault-tolerant gates offer an impressive improvement in fidelity, the computation as a whole is not below the fault-tolerance threshold because of noise associated with state preparation and measurement on this device.
\end{abstract}

\maketitle

We have entered an exciting stage in the development of quantum computers. 
Small scale, prototype quantum devices with a limited number of qubits are beginning to appear, and companies such as IBM and Rigetti are making such devices available in the cloud. 
Although current quantum devices tend to be too small, have limited interconnectivity between qubits and are too noisy to allow meaningful quantum computation, they are an important step forward in the aim to build a large scale universal quantum computer. 
These small devices are sufficient to act as the test bed for proof of principle concepts such as the implementation of simplified quantum algorithms~\cite{coles2018}, quantum walks~\cite{balu2017}, quantum machine learning~\cite{Rist2017} and testing the ability to detect and correct errors. 

Codes that can be implemented on current small noisy quantum devices are perfect testbeds for the ideas of fault tolerance. 
The question we address here is whether such a code can conclusively show a benefit for encoded computation on current small scale noisy quantum devices.
As we discuss later, it is not obvious that the type of noise that exists in current quantum devices will be amenable to such codes.
We study the \ftt~error detecting code, which is one of the smallest interesting codes. 
For instance, it can be concatenated with the toric code~\cite{brell2011,criger2016}, can be viewed as one of the faces of the distance 3 color code~\cite{Bombin2006}, or alternatively as an encoding layer of the C4/C6 code of Ref.~\cite{Knill2005}. 

Experimental quantum error correction and fault tolerance is still in its infancy, but several impressive results have been achieved. Ref.~\cite{Linke2017} demonstrates the ability to prepare logical qubits of the \ftt~code fault tolerantly using trapped atomic ions and Ref.~\cite{Takita2017} replicates this on superconducting qubits. 
Refs.~\cite{Kelly2015etal, wootton2017} implement repetition codes and observe protection against bit-flip errors. 
Ref.~\cite{Ofek2016etal} uses quantum error correction to extend the lifetime of quantum information stored in a superconducting resonator, and Refs.~\cite{Barends2014etal, Heeres2017} implement logical gates inside of a quantum code. 

Recently, \citet{gottesman2016} suggested that the \ftt~code could be implemented fault tolerantly with only five physical qubits. 
He argued that for a small experiment to conclusively demonstrate fault tolerance, the following must be met:
(1)~The encoded circuit must have a lower error rate for all circuits in the family of circuits of interest.
(2)~They must be complete circuits, i.e.\ they must include the initial state preparation and the final measurements.
(3)~The original circuit and the encoded circuit must, in the absence of error, produce the same output distribution.
(4)~It is only meaningful to compare error rates between circuits implemented in the same system.
Additionally, Gottesman noted that for the purposes of demonstrating fault tolerance it is still interesting to consider circuits created from non-universal gate sets.

\citet{vuillot2017} attempted to meet Gottesman's criteria using the \ftt~code on the IBM Quantum Experience (\textbf{IQX}). 
Vuillot comprehensively explored the different types of circuits available in the \ftt~code, though the methodology used meant it was difficult to extract a clear signal from the data as to the overall benefit (or lack thereof) in utilizing the \ftt~code on that noisy device. 
More recently \citet{willsch2018} simulated the \ftt~code numerically and tested it on the IQX. 
Their analysis suggests that a fault tolerant protocol could improve performance provided that errors are due to control errors rather than dominated by decoherence. 

While the inclusion of state preparation and measurement (\textbf{SPAM}) errors in Gottesman's proposed protocol are necessary steps in demonstrating that a computer might meet the \textit{fault tolerance threshold}, the inclusion of such errors is not necessary in demonstrating that the use of fault-tolerant gates and appropriate logical encoding improve the fidelity of the logical gates compared to their physical counterparts. 
In particular, an improvement in the fidelity of such logical gates will address whether the theoretical benefits expected from encoding quantum information can be physically realized or if the noise profile (such as correlated errors) will preclude any such benefit.
The answer to this question is not obvious.
Consequently, rather than examining the error rate of particular circuits created from fault-tolerant gates, we measure the infidelity of the Clifford group elements created from such gates and the circuits arising from such Clifford elements. 
Given the proposed uses of the \ftt~code discussed above, the ability to accurately create such gates is of high interest. 
We look at the average infidelity of the gates (defined later) and demonstrate a conclusive advantage to using such fault tolerant gates to generate these elements and circuits. 
We do this by utilizing the technique of randomized benchmarking (\textbf{RB})~\cite{Emerson2005, Knill2008} to measure the gate errors in a way that is robust to SPAM errors to show that the logical two-qubit gates outperform their physical counterparts. 
Using RB in the logical code space~\cite{lrb2017} of the \ftt~code, we demonstrate that the two qubit average infidelity decreases from $5.8(2)\%$ to $0.60(3)\%$, an improvement of roughly an order of magnitude. 

However, our results do \emph{not} imply that the IBM Q R\"uschlikon device is operating below the fault tolerance threshold. 
As emphasized by \citet{gottesman2016}, the threshold involves improving all aspects of a quantum circuit after the encoding, including the SPAM errors, and our approach using RB is insensitive to SPAM. 
Thus, while the fault-tolerant encoded gates do improve gate fidelity, the IQX device is not yet convincingly below threshold for complete circuits.

\paragraph{Background.}
RB provides an efficient method for the partial characterization of the quality of a gate implementation in a way that is robust to SPAM and to small, arbitrary gate-dependent noise~\cite{wallman2018, merkel2018}. 
It gives the fidelity between the identity channel and the averaged noise $\mathcal{E}$ on the gate set. 
The (Haar) \textbf{average fidelity} is 
\begin{equation}
F(\mathcal{E}) = \int \mathrm{d}\psi \bra{\psi}\mathcal{E}\bigl(\ket{\psi}\!\bra{\psi}\bigr)\ket{\psi}\,,\label{eq:fidelity}
\end{equation} 
and the \textbf{average infidelity} is $1-F$. 

Recent work by \citet{cd2018} relates average fidelity to the \textit{gate set circuit fidelity}, a quantity which compares all possible sequences of circuits of $m$ noisy gates implemented from the noisy gate set $\tilde{\mathbb{G}}$ to their ideal analog in $\mathbb{G}$. 
A \textit{circuit} is a sequence of $m$ elementary gates, and the average gate set circuit fidelity is $\mathcal{F}(\tilde{\mathbb{G}},\mathbb{G},m):=\mathbb{E}[F(\tilde{\mathcal{G}}_{m:1},\mathcal{G}_{m:1})]$, where $\tilde{\mathcal{G}}$ and $\mathcal{G}$ are noisy and ideal gates drawn from the respective gate sets. 
Ref.~\cite{cd2018} proves for a single qubit (and conjectures, with numerical evidence, for two qubits) that, other than a potential SPAM mismatch, the two fidelities are closely related.
This confirms the average infidelity of a gate set as an appropriate metric even when one is considering the fidelity of circuits built from such gates.

\paragraph{The 4 Qubit Code.}
The \ftt~code is defined by the stabilizer generators $XXXX$ and $ZZZZ$~\cite{Shor1995, Bacon2006}. 
Ref.~\cite{gottesman2016} details how to implement it in a fault tolerant manner on a system with limited connectivity. 
To measure the logical qubits in the computational basis, one simply measures all four qubits.
Odd parity heralds an error and the run should be discarded. 

There are 8 logical gates that make up the \textbf{code gate set}.
They split into 6 `active' gates and two `virtual' gates that can be implemented in software by relabeling the physical qubits.
These are shown in the table below, where $P$ is the phase gate ($\mathrm{diag}(1,i)$), SWAP$_{ij}$ swaps qubits $i$ and $j$, \mbox{C-Z} is the controlled-Z gate, and CNOT$_{ij}$ acts from control $i$ to target $j$. 

%\begin{figure}
\begin{center}
\small
\begin{tabular}{m{4cm} m{3cm}}
\toprule
\textbf{Physical Gates}&\textbf{Logical Equivalent}\\ \midrule
$X^{\otimes 4}, Z^{\otimes 4}$ & $I\otimes I$\\
$X \otimes I \otimes X \otimes I$ & $X\otimes I$\\
$X \otimes X \otimes I \otimes I$ & $I\otimes X$\\
$Z \otimes Z \otimes I \otimes I$ & $Z\otimes I$\\
$Z \otimes I \otimes Z \otimes I$ & $I\otimes Z$\\
$H \otimes H \otimes H \otimes H$ & SWAP$_{12} \circ (H\otimes H)$\\
$P \otimes P \otimes P \otimes P$ & $(Z \otimes Z)\,\circ\,$(C-Z)\\ \midrule
SWAP$_{12}$ & CNOT$_{12}$ \\
SWAP$_{13}$ & CNOT$_{21}$ \\
\bottomrule
\end{tabular}
%\caption{Mapping between logical and physical gates for the [4,2,2] code}\label{logicalgates}\label{table:gates}
\end{center}
%\end{figure}

The final ingredient is the logical $\ket{\overline{00}}$ state preparation. 
Although Ref.~\cite{gottesman2016} suggests a method for doing this fault tolerantly, the architecture of the IBM-QX5 means implementation is costly in terms of gates required. 
Since RB is robust to SPAM it was decided not to prepare the $\ket{\overline{00}}$ state in a fault tolerant manner (see~\cite{supplement} for a fuller discussion).

\paragraph{RB with the Real Clifford Group.}
RB uses long sequences of gates with the aim of amplifying small errors in the implementation of these gates. 
By choosing the sequences of gates from a unitary 2-design~\cite{Dankert2009, Gross2007a} the average noise channel $\mathcal{E}$ over random sequences of such gates reduces to a depolarizing channel with the same average fidelity $F(\mathcal{E})$. 

By sampling over a unitary 2-design, the integral in \autoref{eq:fidelity} is replaced by a sum over the design. 
Often the Clifford group is chosen as the unitary 2-design, however in this case the phase gate cannot be implemented in the \ftt~code in a fault tolerant manner. 
If we limit ourselves to fault tolerant gates, the lack of a phase gate makes the Clifford group inaccessible.

Our results make essential use of a modified version of RB called Real RB~\cite{brown2018,Hashagen2018}. 
In Real RB, one can relax the unitary 2-design condition to any \textit{orthogonal} 2-design and still effectively perform RB with the following protocol:
\begin{enumerate}
\item choose a sequence length $m$ and prepare a state ($\rho$), traditionally in the computational basis (but see later), 
\item apply a chosen number of random gates $m$, independently and uniformly drawn from the orthogonal 2-design followed by a further \textit{inversion gate}, which ideally would result in a net sequence equal to the identity,
\item measure $\rho$ with E (an effect operator of a POVM), to determine if it has been returned to the starting state.
\end{enumerate}
The above steps are repeated in order to estimate the survival probability $\bar{q}$ over a range of sequence lengths.
Ref.~\cite{Hashagen2018} proves that this can be fit to the model
\begin{equation}\label{eq:realModel}
\bar{q}(m,E,\rho)=A + b^mB + c^mC\,,
\end{equation}
where, with $\rho_\pm = \tfrac{1}{2}(\rho\pm\rho^T)$, the constants $A$, $B$, and $C$ are
\begin{align}
A &= \tfrac{1}{d}\text{Tr}[E], & 
B &= \text{Tr}\bigl[E\rho_+\bigr] - A, & 
C &= \text{Tr}\bigl[E \rho_-\bigr].
\end{align}
For two qubits we also have $F(\mathcal{E}) = (9b + 6c + 5)/20$.

The code gate set generates a subgroup of $\mathsf{C}(2)$, the two-qubit Clifford group, that we call the \textbf{Realizable Group}, $\mathsf{R}(2)$. 
The group $\mathsf{R}(2)$ has only 576 elements, in contrast to the 11,520 elements of $\mathsf{C}(2)$. 
Moreover, the average number of elementary gates in a group element is reduced from about 7 in $\mathsf{C}(2)$ to only about 4 in $\mathsf{R}(2)$, making it more efficient for RB as well.
Finally, $\mathsf{R}(2)$ provably yields an orthogonal 2-design, so the Real RB protocol applies~\cite{supplement}.
There are subtleties that may impact when fitting the above formulas to an error detecting experiment, although these did not impact our analysis.
For a fuller discussion see~\cite{supplement}.

\paragraph{Implementation Details.} 

We implemented logical RB and physical RB on the IQX using $\mathsf{R}(2)$ as the twirling group to measure the average infidelities. 
If there is an advantage to computing in the logical space defined by the \ftt~code, then the average infidelity will be lower.

In the supplementary material~\cite{supplement} we detail how we perform RB in such a way as to eliminate the fitting parameter $A$. In addition, for certain obvious choices of $\rho$, $B$ or $C$ becomes zero~\cite{Hashagen2018}. 
By choosing an appropriate $\rho$ we can then fit the sequences from those experiments to a simpler fitting sequence. 
In the computational basis $\rho=\ket{00}\bra{00}$ and $\rho_-=0$, thus $C$=0. 
Consequently if we perform the RB protocol discussed in~\cite{supplement} with $\rho = \ket{00}\bra{00}$, we can fit to the simplified formula $\bar{q}(m,\ket{00}\bra{00})=Bb^m+0.25$, to determine $b$. 
This is similar to the ideas discussed in Refs.~\cite{Dugas2015, Muhonen2015etal, Fogarty2015}, and yields improved parameter estimates for the same sample size. 

%------------------------------
\begin{figure}
\includegraphics[width=.95\columnwidth]{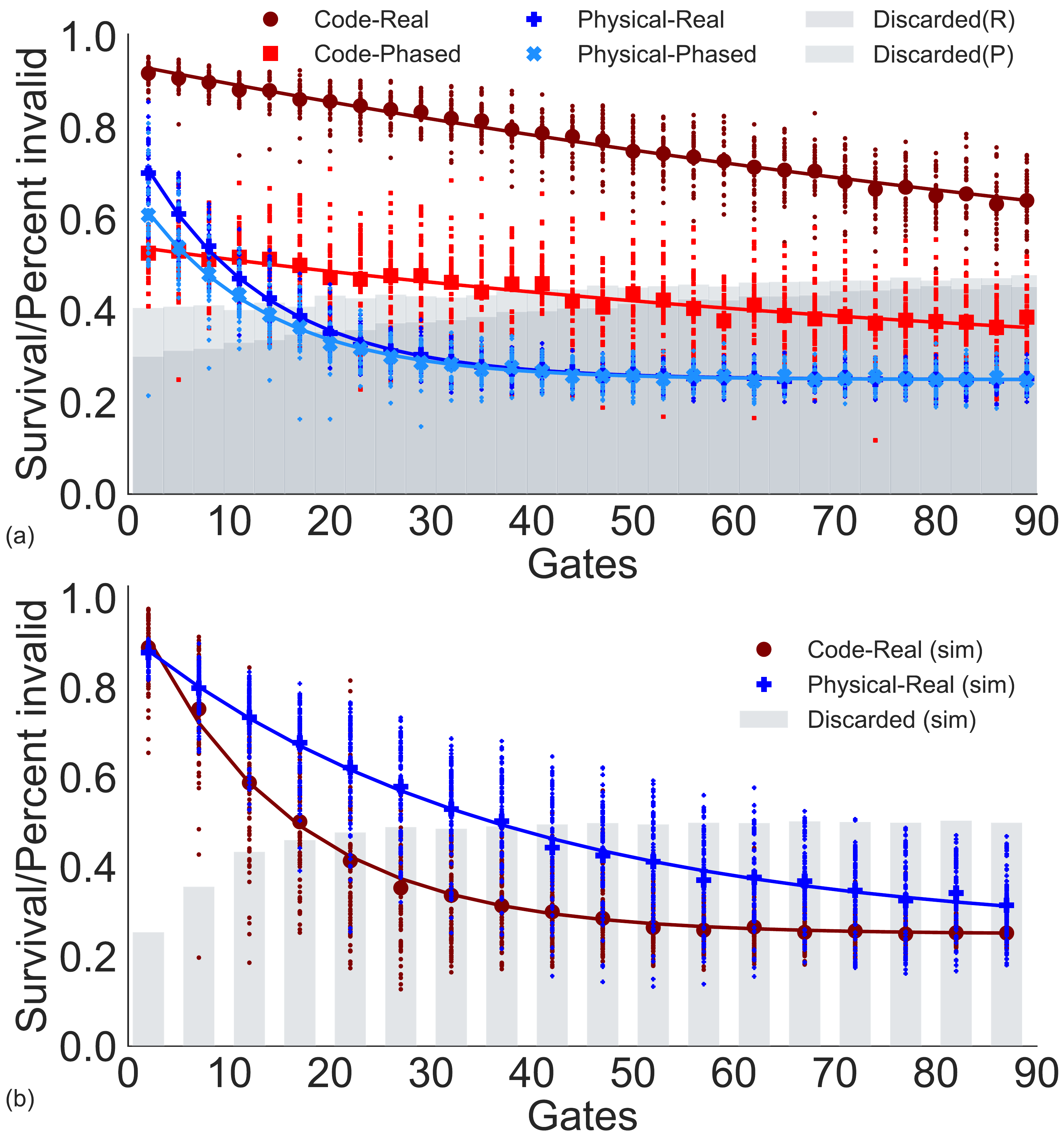}
\caption{(a) Fitting charts for the four different types of runs, used to extract the $b$ and $c$ parameters in \autoref{eq:realModel}. The data was fit as described in the supplementary material~\cite{supplement}.
Post-selection on invalid code states was used to throw away runs where an error was detected. 
The bars show the percentage of detected (and discarded) runs for the \ftt~code runs. 
The number of initial errors are greater in the phase run because of the need to perform an initial non-fault tolerant gate (see text).
(b) Simulation showing survival probabilities for logical `real-Clifford' RB using the  \ftt~code and the equivalent physical 2-qubit RB.
(Here we have only shown the `real' runs, not the `phased' runs.)
The same high fidelity 4-qubit noise map (fidelity $\approx 0.99$) is used for the runs although additional noise is applied to the CNOTs of the run on physical qubits. The noise map contains single-qubit errors as well as $ZZ$ errors between neighboring qubits.
Although the combined noise map is high fidelity, the correlated errors are sufficiently strong that the fidelity \textit{decreases} when using encoded gates, meaning the encoded computation performs worse than the bare physical computation in the correlated noise scenario.
}\label{fig:fitting}\label{fig:ZZcode}
\end{figure}
%------------------------------

To determine the $c$ parameter we need to move outside the computational basis for our choice of $\rho$. 
In this case the optimal method is to rotate~$\rho$ by applying a phase gate followed by a Hadamard, inverting this after the twirl and measuring $\bra{00}$ in the normal way. 
Since a fault tolerant phase gate cannot be applied in the code space, to measure $c$ we need to use a non-fault tolerant phase gate together with the encoded Hadamard; see \cite{supplement} for details. 
When working outside the code, we can implement the phase and Hadamard gates directly.

In total, this fit procedure requires that we perform \textbf{four types of runs} for the \ftt~code: (1) the run in the computational basis (a \textbf{standard run}) and (2) the run initialized with the non-fault tolerant phase gate (a \textbf{phased run}); and for the physical qubits (3) the standard run in the computational basis and (4) the phased run.

\paragraph{Analysis.}
The four averaged fitting charts are shown in \autoref{fig:fitting}(a). 
This allows us to extract values for $b$ and $c$ and calculate the fidelity/infidelity using \autoref{eq:fidelity}. 
The calculations yield an infidelity of $0.60(3)\%$ for the \ftt~code and $5.8(2)\%$ for the 2 physical qubits. 
The uncertainties were obtained through bootstrapping the experimental data~\cite{supplement}. 

The question then remains: how much of the benefit is derived from the ability of the code to detect errors and how much is derived from the virtual CNOTs once computation has been moved into the code space?

We can use the arguments presented in Ref.~\cite{lrb2017} to determine this. 
If, instead of using post-selection to discard runs that we know are in an erroneous code state, we fit to the complete set of runs obtained for the \ftt~code, then this gives us a method of comparing the performance of the code where we are detecting errors and one where we are not.
With the caveats that these runs were conducted at different times from the main runs and that a naive RB fit to the data was used (see \cite{supplement}), this formulation gave an infidelity that was less than $5.8\%$, but still significantly higher than the infidelity with error detection. 
This confirms two things. 
First, there is indeed a benefit in moving computation into a code that allows virtualization of the CNOTs, and secondly, there is still an additional substantial benefit to using the code to detect errors.

\paragraph{Comparison to correlated noise.} 
Data on the quantum devices made available by IBM show that there is significant crosstalk between connected qubits.
One of the interesting questions answered by this paper is whether such crosstalk causes correlations strong enough to defeat any fidelity gains that might occur from using encoded gates operated in a fault tolerant way.
To illustrate why this might be the case we note that a natural noise model for such correlated noise is to model it as noise of the form $e^{i\theta Z_i Z_{i+1}}$ for small $\theta$. This model accounts for some of the correlated noise observed in recent ion trap experiments~\cite{Nigg2014} and analyzed in Ref.~\cite{Robertson2017}.

In \autoref{fig:ZZcode}(b) we show, by simulation, that even with a high fidelity noise map, the correlated noise can be sufficient to prevent any benefit being observed from computing in the logical code space of the \ftt~code.

\paragraph{Discussion.}
A useful error detecting code consists of 1) a code space in which logical quantum gates can be performed on the encoded quantum information; 2) an ability to perform certain fault-tolerant gates within the code space; and 3) the ability to detect if a certain number of errors (limited by the distance of the code) have occurred. 
We have demonstrated how variations of the RB protocol~\cite{Hashagen2018,lrb2017} can be used to determine, in a well defined and principled manner, whether the fault-tolerant quantum gates that are supported by such a code space can be performed with lower infidelity than such equivalent gates in the raw physical qubits. 
These specific RB variations were essential since the \ftt~code is not able to perform the full Clifford group fault tolerantly. 

The substantial decrease in the infidelity we observed appears to come from two sources. 
The first is that it virtualizes the noisiest gate, the CNOT gate, resulting in a decrease in the infidelity of the averaged gate set noise. 
The second is that by allowing error detection (and subsequent post-selection) we see a further decrease in the infidelity of the averaged gate set, leading to an overall decrease in infidelity of the fault tolerant gate set by a factor of 10. 

We learn from these experiments that the IBM device does not have noise correlations that are so strong as to preclude an improved fidelity when using an encoded gate.
As we showed via numerical simulations (\autoref{fig:ZZcode}(a)) a natural noise model~\cite{Nigg2014, Robertson2017} involving 2-qubit correlated errors between neighbors \textit{can} preclude an improved fidelity even when the average error rates are comparable to those we observe in the IBM device.
More experiments are required to fully understand the role that noise correlations play in this device in the context of fault tolerance.

The analysis of Ref.~\cite{cd2018} confirms that the average fidelity as measured by RB is closely related (as particularized in their analysis) to the average fidelity for all relevant circuits. 
Given this, it is clear that the average gate set fidelity will be substantially higher if the fault tolerant gates are used, and we argue that this suffices for a conclusive demonstration of the benefit of using an error detecting code with fault tolerant gates.
Recent work has also shown~\cite{Huang2018, Beale2018} that encoded noise tends to be less coherent than unencoded noise.
Consequently, an improvement in average gate fidelity is also likely to herald an improvement in other metrics relevant for fault tolerance such as the diamond distance, although more experiments would be required to establish this conclusively.

There is at present no complete theory for fitting an RB decay curve with post-selection.
While this had minimal impact on our present study (for the reasons discussed in~\cite{supplement}) developing a full theory will be essential for more detailed studies.
A further exciting step would be to employ the techniques used here together with those discussed in Ref.~\cite{lrb2017} to benchmark error-corrected gates. 
While devices with sufficient connections to support one, or maybe a few, error-corrected qubits are on the verge of becoming available for general experimentation, current implementations of IQX do not yet have conditional measurements or reinitialization. 
In the meantime, error detecting codes on more qubits and implementing different fault-tolerant gate sets can be examined using the techniques discussed above.

%------------------------------
\begin{acknowledgments}
We thank D.~Gottesman for comments on our draft and T.~Yoder for highlighting the impact of post-selection on the RB decay curve.
This work was supported by the US Army Research Office grant numbers W911NF-14-1-0098 and W911NF-14-1-0103, and by the Australian Research Council Centre of Excellence for Engineered Quantum Systems CE170100009. 
\end{acknowledgments}

\title{Supplementary Material: Fault-Tolerant Logical Gates in the IBM Quantum Experience}
\author{Robin Harper}
\affiliation{Centre for Engineered Quantum Systems, School of Physics, The 
University of Sydney, Sydney, Australia}
\author{Steven Flammia}
\affiliation{Centre for Engineered Quantum Systems, School of Physics, The 
University of Sydney, Sydney, Australia}
\affiliation{Yale Quantum Institute, Yale University, New Haven, CT 06520, USA}
\date{\today}
\appendix*
%\maketitle

\vspace{-10pt}
\section{Supplementary material}

\paragraph{Details of implementation.} 

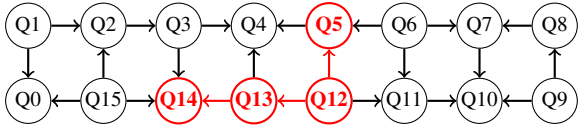
\begin{figure}
{\footnotesize \begin{tikzpicture}
		\node[circle,draw=black, fill=white,inner sep=0pt,minimum width=6mm] (v1) at (0,1) {Q1};
		\node[circle,draw=black, fill=white,inner sep=0pt,minimum width=6mm] (v2) at (1,1) {Q2};
		\node[circle,draw=black, fill=white,inner sep=0pt,minimum width=6mm] (v3) at (2,1) {Q3};
		\node[circle,draw=black, fill=white,inner sep=0pt,minimum width=6mm] (v4) at (3,1) {Q4};
		\node[circle,thick,draw=red, fill=white,inner sep=0pt,minimum width=6mm] (v5) at (4,1) {\textcolor{red}{\textbf{Q5}}};
		\node[circle,draw=black, fill=white,inner sep=0pt,minimum width=6mm] (v6) at (5,1) {Q6};
		\node[circle,draw=black, fill=white,inner sep=0pt,minimum width=6mm] (v7) at (6,1) {Q7};
		\node[circle,draw=black, fill=white,inner sep=0pt,minimum width=6mm] (v8) at (7,1) {Q8};
		\node[circle,draw=black, fill=white,inner sep=0pt,minimum width=6mm] (v0) at (0,0) {Q0};
		\node[circle,draw=black, fill=white,inner sep=0pt,minimum width=6mm] (v15) at (1,0) {Q15};
		\node[circle,thick,draw=red, fill=white,inner sep=0pt,minimum width=6mm] (v14) at (2,0) {\textcolor{red}{\textbf{Q14}}};
		\node[circle,thick,draw=red, fill=white,inner sep=0pt,minimum width=6mm] (v13) at (3,0) {\textcolor{red}{\textbf{Q13}}};
		\node[circle,thick,draw=red, fill=white,inner sep=0pt,minimum width=6mm] (v12) at (4,0) {\textcolor{red}{\textbf{Q12}}};
		\node[circle,draw=black, fill=white,inner sep=0pt,minimum width=6mm] (v11) at (5,0) {Q11};
		\node[circle,draw=black, fill=white,inner sep=0pt,minimum width=6mm] (v10) at (6,0) {Q10};
		\node[circle,draw=black, fill=white,inner sep=0pt,minimum width=6mm] (v9) at (7,0) {Q9};
		\draw[thick,->] (v1) -- (v2);
		\draw[thick,->] (v2) -- (v3);
		\draw[thick,->] (v3) -- (v4);
		\draw[thick,->] (v5) -- (v4);
		\draw[thick,->] (v6) -- (v5);
		\draw[thick,->] (v6) -- (v7);
		\draw[thick,->] (v8) -- (v7);
		\draw[thick,->] (v9) -- (v10);
		\draw[thick,->] (v11) -- (v10);
		\draw[thick,->] (v12) -- (v11);
		\draw[thick,->,draw=red] (v12) -- (v13);
		\draw[thick,->,draw=red] (v13) -- (v14);
		\draw[thick,->] (v15) -- (v14);
		\draw[thick,->] (v15) -- (v0);
		\draw[thick,->] (v1) -- (v0);
		\draw[thick,->] (v15) -- (v2);
		\draw[thick,->] (v3) -- (v14);
		\draw[thick,->] (v13) -- (v4);
		\draw[thick,->,draw=red] (v12) -- (v5);
		\draw[thick,->] (v6) -- (v11);
		\draw[thick,->] (v7) -- (v10);
		\draw[thick,->] (v9) -- (v8);
\end{tikzpicture}}\\
\caption{Qubit connectivity diagram in the IBM-QX5 chip. 
The marked qubits were used to implement the \ftt~code.}
\label{fig:topology}
\end{figure}

The IBM Q Experience (\textbf{IQX}) is an initiative by IBM that allows access to some of IBM's quantum devices.
The experiments reported here were carried out on the IBM-QX5, a 16 qubit device with connectivity as shown in \autoref{fig:topology}. 
Details about the device can be found on the IBM website \footnote{{https://quantumexperience.ng.bluemix.net/qx/devices}}. 

The interface to the IBM-QX5 is through a quantum assembly language QASM~\cite{2017arXiv170703429C}. 
Typically methods of error correction rely on measuring qubits and then acting on the measurement (or, in the case of measurement-free error correction, being able to reinitialize the qubits). 
The elements of QASM required to do this have not yet been implemented on the devices provided by IBM and, accordingly, there is no obvious way to perform error correction though the interface provided. 
Nevertheless, QASM suffices for running an error detecting code with post-selection. 
That is, at the end of the run the measurements allow (in the case of the \ftt~code) the detection of a single (or potentially odd-number) of errors, so that those runs known to be erroneous can be discarded. 
Although this results in fewer successful runs, we show that the runs that do succeed have a substantially improved fidelity.

For the experiments on the `raw' physical qubits, qubits 12 and 5 were chosen as the physical qubits.
They form part of the logical qubit array allowing valid comparisons between the regimes. 
(A different qubit pair was also tested with similar results --- see below.)

The QASM specification allows sequences of gates to be compiled into shorter composite gates unless a `barrier' command is inserted. 
Barrier commands were only inserted between discrete group elements of $\mathsf{R}(2)$. 
Since the QASM compiler would attempt to shorten the number of gates required for each element of the group, all infidelities are only reported on a `per real Clifford' basis.

The quantum chip on the IQX has a life cycle that is relevant to any comparison runs. 
Twice a day the machine is taken offline and re-calibrated. 
Fidelities on the machine tend to be higher shortly after a re-calibration cycle and worse just before one. 
Also the machine is not sole use, so multiple jobs might be separated in time depending on how many other users are submitting jobs. 
Runs can be batched, but there is an overall QASM job size, which means that in order to gain useful statistics multiple jobs had to be submitted.

For our purposes, the comparative infidelity between logical and physical runs is more important than the absolute infidelities. 
Given this, job submissions were delayed until low use times and then batched so that the different types of run sequences were executed as close to each other in time as possible. 
A combination of patience and an advantageous time zone meant that the data reported here were largely gathered over a period of three maintenance cycles, with the different types of runs interleaved throughout the data gathering period.

We refer to each batch of runs carried out on the IQX as an experiment. 
An experiment consisted of multiple sequences of random sets of gates starting at 2 Clifford gates (followed by an inverting gate), and incrementing by 3 until 92 real Clifford gates had been reached (followed by an inverting gate). 
That is, each experiment consisted of a total of 30 random sequences each with a different gate length. 
Every sequence was measured 1024 times to build up the appropriate statistics. 

Each experiment was performed 4 different ways, each time the sequences being randomized and a different ($II$, $IX$, $XI$ or $XX$) being compiled into the sequence. 
For a standard run this was compiled in at the beginning of the sequence, and for a phased run it was compiled in after the state had been returned to the computational basis. 
This was done for each of the four types of run. 
Finally all of the above was repeated 9 times, meaning we had a total of 36 sequences at every gate length for each of the four types of run.

\paragraph{Logical $\ket{\overline{00}}$ state.} Gottesman~\cite{gottesman2016} proposed that the logical $\ket{\overline{00}}$ state could be prepared fault tolerantly using the circuit shown in \autoref{fig:statePrep}(b). 
The ancilla qubit is used to detect a (single) error in the state preparation. 
If it is measured as `1' then the state has not been properly prepared. 
With the IBM-QX5 there is no ring of 5 qubits, which is the architecture proposed in~\cite{gottesman2016}. 
It is still possible to perform the circuit illustrated in \autoref{fig:statePrep}(b), but there is some additional complexity involved. 
Using \autoref{fig:topology} as a reference, one can see that performing a CNOT from Q14$\rightarrow$Q3 (this involves 4 Hadamards and a CNOT), a swap of qubits 3 and 4 (requiring 4 Hadamards and 3 CNOT gates) and finally a CNOT from Q5$\rightarrow$Q4 we have an equivalent circuit, at the cost of an additional 8 single qubit and 5 two-qubit gates. 
(A similar gate count is required if one uses the following identity $\text{CNOT}_{13} = \text{CNOT}_{23}\text{CNOT}_{12}\text{CNOT}_{23}\text{CNOT}_{12}$.)

Some initial testing showed that the success rate of state preparation without the ancilla was comparable to (and often better than) the success rate of state preparation with the ancilla (and post-selection based on the ancilla measurement). 
Since RB is robust to state preparation errors (and measurement errors), the need to prepare the state in a fault tolerant way was not something we needed to do in order to determine if fault tolerant computation in the code space was improved as compared to raw computation in the physical qubit space. 
For those reasons, all the results in this paper were done with the straightforward state preparation (i.e.\ only the first four gates in \autoref{fig:statePrep}(b)). 
In practice, with a custom architecture, it is likely that fault tolerant state preparation would be preferable.

\paragraph{Real RB.} The real Clifford group $\mathsf{C}_\mathbb{R}(n)$ on $n$ qubits is the normalizer in the orthogonal group $\mathsf{O}(2^n)$ of the real Pauli group $\mathsf{P}_\mathbb{R}(n)$, 
\begin{align}
\mathsf{C}_\mathbb{R}(n) \coloneqq \{ O \in \mathsf{O}(2^n)|O \mathsf{P}_\mathbb{R}(n) = \mathsf{P}_\mathbb{R}(n)O\}\,.
\end{align}
We recover the full Clifford group $\mathsf{C}(n)$ by replacing the orthogonal group with the unitary group $\mathsf{U}(2^n)$ and the real Pauli group with the full Pauli group $\mathsf{P}(n)$. 
Here it is understood that the group always carries with it the standard unitary representation on $n$-qubits as a subgroup of $\mathsf{C}(n)$, the full Clifford group. 
The Realizable Group $\mathsf{R}(2)$ discussed in the main text is a subgroup of $\mathsf{C}_\mathbb{R}(2)$.

The frame potential~\cite{Gross2007a} of a unitary representation of a group $\mathsf{G}$ is defined as:
\begin{equation}
\mathcal{P}(\mathsf{G}) = \frac{1}{|\mathsf{G}|}\sum\limits_{g\in \mathsf{G}}\lvert\text{Tr}(g)\rvert^4.
\end{equation}
Ref.~\cite{Hashagen2018} proved the following: 
\begin{enumerate}
\item The group $\mathsf{C}_\mathbb{R}(n)$ in the standard representation forms an orthogonal 2-design; 
\item A group $\mathsf{G}$ is an an orthogonal 2-design if and only if $\mathcal{P}(\mathsf{G}) = 3$; 
\item An orthogonal 2-design can be used as the twirling group in RB with the protocol listed in the main text.
\end{enumerate}
From points 2 and 3, we only need to show that $\mathcal{P}\bigl(\mathsf{R}(2)\bigr) = 3$ and it follows that we are justified in using the Real RB protocol with the Realizable Group. 
This can be verified by explicit computation using a computer algebra package. 

The full two-qubit Clifford group $\mathsf{C}(2)$ consists of 11,520 distinct elements. 
The real Clifford group $\mathsf{C}_\mathbb{R}(2)$ on two qubits is smaller, with only 1,152 elements. 
The Realizable Group $\mathsf{R}(2)$ has 576 elements, so this is 20 times smaller than the full Clifford group and half as large as the real Clifford group. 
Assuming generators given by the X, Z, phase, Hadamard, and a CNOT gate, the average number of these basic gates required to implement $\mathsf{C}(2)$ is just over 7, and this remains true even if the alternate Clifford unitary 2-design described in Ref.~\cite{Cleve2015} is used (which has only 960 elements). 
By contrast $\mathsf{R}(2)$ has on average only just over 4 generators.

\begin{figure}[t!]
\[\qquad \begin{array}{c}
\Qcircuit @C=.6em @R=.6em {
  \lstick{|q_1\rangle}	& \gate{H}	& \targ 	& \gate{P}	& \targ 	& \qw \\
  \lstick{|q_2\rangle}	& \gate{H}	& \ctrl{-1} 	& \qw 	& \ctrl{-1}	& \qw \\
  \lstick{|q_3\rangle}	& \gate{H}	& \qw 	& \qw 	& \qw 	& \qw \\
  \lstick{|q_4\rangle}	& \gate{H}	& \qw 	& \qw 	& \qw 	& \qw 
}\end{array}
 \ \equiv \qquad\ \ 
\begin{array}{c}
\Qcircuit @C=1em @R=.8em {
  \lstick{|L_1\rangle}	& \gate{H}	& \qswap		& \gate{P}	& \qw \\
  \lstick{|L_2\rangle}	& \gate{H}	& \qswap\qwx 	& \qw	& \qw 
}\end{array}
\]
\[\text{\ (a)\qquad\qquad \ftt~code \qquad\qquad\qquad Logical equivalent\qquad}\]\\[.4em]
\[\Qcircuit @C=1em @R=1em {
  \lstick{|q_1\rangle}	& \qw	& \targ 	& \qw 	& \qw 	& \ctrl{4}	& \qw 	& \qw \\
  \lstick{|q_2\rangle}	& \gate{H}	& \ctrl{-1} 	& \ctrl{1} 	& \qw 	& \qw	& \qw 	& \qw \\
  \lstick{|q_3\rangle}	& \qw	& \qw 	& \targ 	& \ctrl{1} 	& \qw	& \qw 	& \qw \\
  \lstick{|q_4\rangle}	& \qw	& \qw 	& \qw 	& \targ 	& \qw	& \ctrl{1} 	& \qw \\
  \lstick{|A\rangle}	& \qw	& \qw 	& \qw 	& \qw	& \targ	& \targ 	& \qw 
}\](b)
\caption{(a) Circuit used to implement a Hadamard gate followed by a phase gate. 
Although the phase gate is not fault tolerant, it is only used at the beginning and end of the RB sequences, so any errors caused by the gate are absorbed into the SPAM errors.
(b) Circuit for preparing the the logical $\ket{\overline{00}}$ for the \ftt~code in a fault tolerant manner~\cite{gottesman2016}. 
The ancilla is used to determine if an error occurred during the preparation of the state. 
With the IBM QX5, the additional number of physical gates required to perform the CNOTs to the ancilla (see text) counteracted the benefit of such error detection in state preparation. 
This coupled with the fact that RB is robust to SPAM noise, meant that we could view state preparation errors as part of the SPAM, not impacting the ability to measure the fidelity of the logical qubits in implementing the real Clifford gates. 
Consequently, only the first four gates were used in state preparation.
}\label{fig:statePrep}
\end{figure}

\begin{figure}
\includegraphics[width=1\columnwidth]{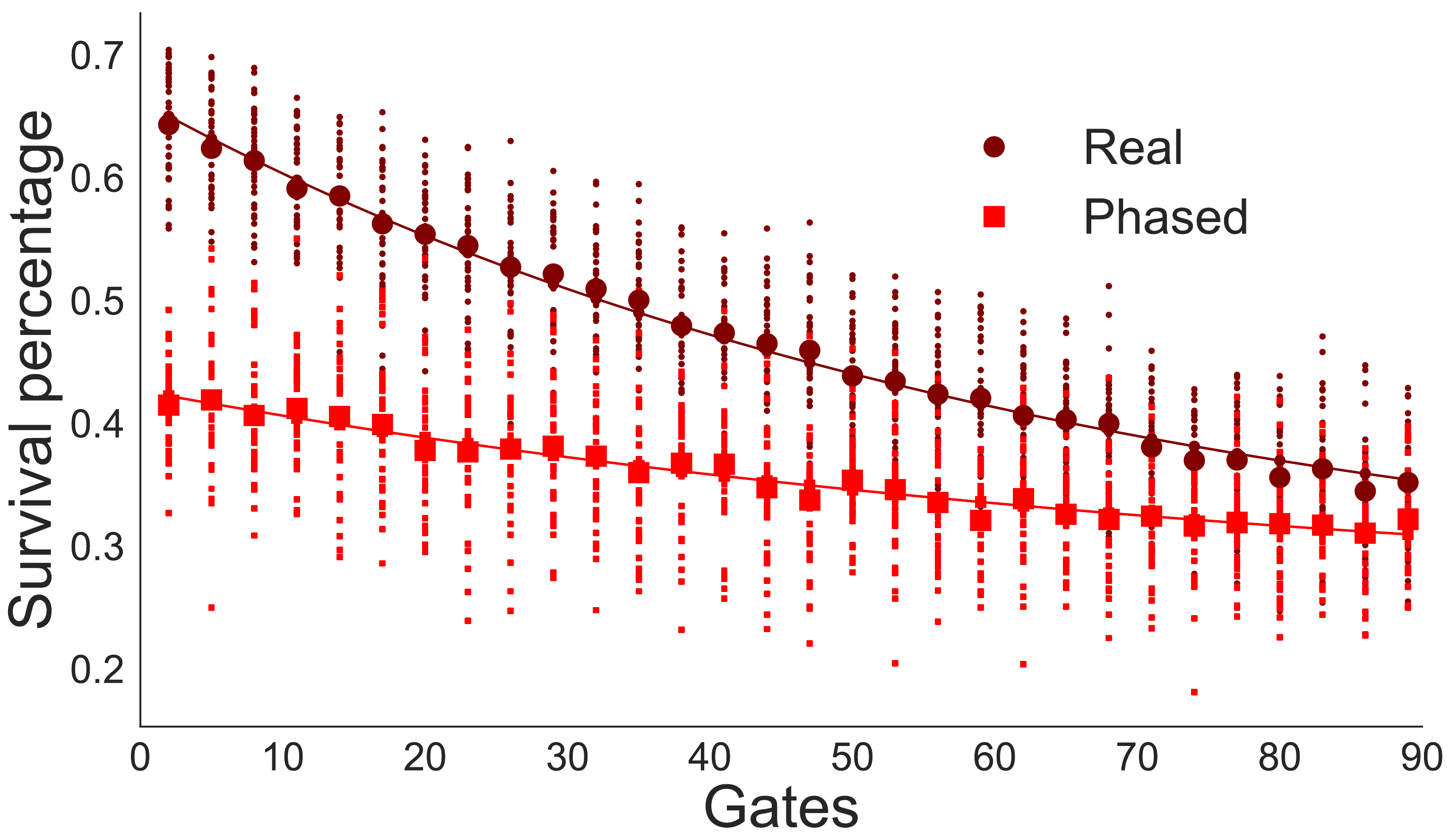}
\caption{Fitting chart for the \ftt~code runs, but without post-selection. 
In this case we do not throw away sequences which end up outside the code space. 
This allows us to approximately quantify the benefit of operating inside the code space (where we can use virtual CNOTs) as compared to the joint benefit of operating inside the code space and being able to detect some errors.
A better fit could be obtained in the \ftt~code runs using longer sequences, but IBM QASM limits were an issue.}\label{fig:rawcode}
\end{figure}

\paragraph{Simplifying the RB fitting procedure.} As noted in the main text, we can eliminate certain fitting parameters, leading to tighter fits for the same amount of data.

In the case of the IQX the measurement operators are predetermined. 
Measurement is in the computational basis, each qubit being measured separately. 
For each sequence to be executed we specify the number of times we wish to run the sequence and receive back the number of times each possible combination was observed. 
For instance, if we are interested in 2 qubits, then we will receive back the number of times a `11', `10', `01' and `00' were recorded (the sum of these numbers equaling the number of runs made). 
We will refer to such measurements as a probability measurement, i.e.\ as a POVM with effects $E_{\uparrow\uparrow}, E_{\uparrow\downarrow}, E_{\downarrow\uparrow}$ and $E_{\downarrow\downarrow}$ respectively where $E_{\uparrow\uparrow} + E_{\uparrow\downarrow} + E_{\downarrow\uparrow} + E_{\downarrow\downarrow} = 1$.

Noting that the fitting constant $A$ from Eq.~2 is independent of the initial state $\rho$, then we can compile in at the beginning (or end) of the sequence one of the following gates: $II,XI,IX,XX$. 
Then, provided the sequences are chosen independently of the gate that has been compiled in, we can utilize the appropriate measurement (e.g.\ if we start in the $\ket{00}$ state and compile in an $XX$ gate, we would use the $E_{\uparrow\uparrow}$ measurement), average the results and set $A=0.25$. 
This is a simple extension of the ideas discussed Refs.~\cite{Knill2008, Dugas2015, Muhonen2015etal, Fogarty2015}). 
Noise from these additional gates is absorbed into the remaining SPAM parameters.
A careful choice of initial state preparation, as discussed in the main text, then allows us to eliminate either $B$ or $C$ so that the net effect is to fit a straightforward single exponential decay, either $B b^m$ or $C c^m$ depending on the run. 
Using this technique reduces the correlation with the nuisance parameters and avoids contaminating the accuracy or precision of the parameter of interest ($b$ or $c$) with uncertainty in the nuisance degrees of freedom. 

\paragraph{Fitting the curves.} 
One of the interesting aspects of using RB techniques for logical benchmarking is that while the gates used in this paper form an orthogonal 2-design in the logical code space, they do not form such a design in the physical 4-qubit space in which they are implemented. 
This is relevant because where we are looking at the probability of survival $\Pr(S)$ (i.e.\ return to the original state) conditioned on number of gates ($m$) and return to the code space ($c$), then we can write:
\begin{equation}
\label{eq:surv}
	\Pr(S|m,c) = \frac{\Pr(S|m)}{\Pr(c|m)}\,.
\end{equation}
The numerator in \autoref{eq:surv} can be modeled in the logical code space using the RB formula, however the denominator requires an examination of the physical 4 qubit system and the effective twirl of the physical realization of the logical gates in the larger Hilbert space. 
Failure to take into account $\Pr(c|m)$ might lead to inaccurate fits. 
While we have not made an attempt to model $\Pr(c|m)$ in this paper, we note that we have measured the actual rates as ($1-\Pr(c|m)$), which are reproduced as bars in Fig.~1(c) of the main text. 
As can be seen, because 1) we did not use a fault tolerant state preparation; and 2) in the phased runs there were additional noisy non-fault tolerant gates applied, $1-\Pr(c|m)$ is relatively flat with increasing $m$ starting off at 0.3 (0.4) in the Real (Phased) run, slowly rising to its asymptote of 0.5. 
We have therefore approximately modeled it as constant and fit a standard RB curve to $\Pr(S|m,c)$. 
Although this is an approximation we note the goodness of fit to the curves and the fact that the effect noted in the paper (the almost an order of magnitude improvement in infidelity) is greater than any difference that might come from a more precisely modeled fit. 
It does, however, remain important future work to calculate the fit equations exactly.

Similar problems arise with the fit to survival probabilities of logical gates which do not throw away the error states (see \autoref{fig:rawcode}). 
Here a fit to the survival probability in the 4 qubit Hilbert space is not justified as the physical implementation of the logical gates in the 4 qubits gates do not form an orthogonal 2-design in the 4 qubit space (they only form such a design in the code subspace). 
In the case where we are not throwing away runs that are outside the code space, we have a leaky subspace. 
Again we have naively fit the runs assuming a logical sub-space and logical gates.
We justify such a fit by noting that we are not attempting to find a specific value for the infidelity of such gates, rather just a confirmation that there is still substantial benefit in detecting invalid states. 
The runs are clearly lower infidelity than the equivalent 2 physical qubit runs, but also are clearly not as good as the full error detecting runs shown in Fig.~1(c) of the main text.
Since these runs were conducted at a different time from the main runs, even with an exact formula we would not be able to make a direct comparison (as the characteristics of the IQX vary appreciably over time).

\paragraph{Calculating uncertainties.} 
To calculate the uncertainties given in the main text we used non-parametric bootstrapping. 
In this case, for each particular gate length, 36 sequences were sampled with replacement. 
For each sequence so sampled the survival probability of the sequence was used as the estimated probability to generate a binomial random variable, sampled and averaged over: a) in the case of the physical 2 qubit system, the number of shots in the original experiment; and b) in the case of the \ftt~code, the number of error-free runs for that sequence. 
A total of 9,999 such data sets were sampled and used to calculate 9,999 re-sampled infidelities. 
The error bars are the 250\textsuperscript{th} and 9,750\textsuperscript{th} of such infidelities.

\paragraph{Determining the gates for the physical qubits.} The minimal gates required to generate the group for the physical qubits were determined by a simple search. 
We assumed generators of a CNOT gate (qubit 12 $\rightarrow$ 5) and the single qubit gates $X,Z,H$ on either qubit and used the search to determine the minimum generator gate sequences to create each element of the Realizable Group $\mathsf{R}(2)$. 

\paragraph{Checking vis-a-vis other physical qubits.} In order to check the validity of the comparisons made in the main text, further runs were carried out utilizing physical qubits 12 and 13 on the device. 
They are not analyzed here as, for reasons discussed in the main text, it was important to have all runs temporally correlated. However we note that the calculated infidelity for qubits 12 \& 13 was slightly lower ($5.4(2)\%$ compared to the results for qubits 12 \& 5 of $5.8(2)\%$), but such a slight increase has no impact on the conclusions we reach.

\end{document}